\documentclass[letterpaper, 10 pt, conference]{ieeeconf}  

\IEEEoverridecommandlockouts                              

\usepackage{graphics} 
\usepackage{epsfig} 
\usepackage{mathptmx} 
\usepackage{times} 
\usepackage{amsmath} 
\usepackage{amssymb}  
\usepackage{stmaryrd}
\usepackage{hyperref}
\usepackage{subcaption}
\usepackage{thmtools}
\usepackage{multirow}
\usepackage[ruled,vlined]{algorithm2e}

\usepackage[pdftex,dvipsnames]{xcolor}
\usepackage[color=SkyBlue!50]{todonotes}

\usepackage{fainekos-macros}
\usepackage{rod-macros}
\DeclareMathOperator{\sign}{sign}

\title{\LARGE \bf
	Time-Robust Control for STL Specifications
}

\author{Al\"ena Rodionova, Lars Lindemann, Manfred Morari and 
	George J. Pappas\ $^\dagger$
\thanks{$^\dagger$ The authors are with the Department of Electrical and Systems Engineering, University of Pennsylvania, Philadelphia PA, USA. 
  {\tt\small\{nellro,\, larsl,\, morari,\, pappasg\}@seas.upenn.edu}.}%
\thanks{This work was supported by AFOSR Assured Autonomy.}
}

\begin{document}

\maketitle
\thispagestyle{empty}
\pagestyle{empty}

\begin{abstract}
We  present  a robust control framework for time-critical systems in which satisfying real-time constraints  robustly is of utmost importance for the safety of the system. Signal Temporal Logic (STL) provides a formal means to express a large variety of real-time constraints over signals and is suited for planning and control purposes as it allows us to reason about  the time robustness of such constraints. The time robustness of STL particularly quantifies the extent to which timing uncertainties can be tolerated without violating real-time specifications. In this paper, we first pose a control problem in which we aim to find an optimal input sequence to a control system that maximizes the time robustness of an STL constraint. We then propose a Mixed Integer Linear Program (MILP) encoding and provide correctness guarantees along with a complexity analysis of the encoding. We also show in two case studies that maximizing STL time robustness allows to account for timing uncertainties of the underlying control system.

\vspace{10pt}
\end{abstract}

\section{INTRODUCTION}

Consider the following three time-critical systems: an air traffic control center sequencing  arriving airplanes, a group of autonomous robots with the goal to maximize the time within each others communication range, and an automated warehouse in which interleaving processes are required to happen in a pre-specified order. These systems, while being fundamentally different in terms of their dynamics and specifications, share the property that not meeting their real-time constraints can compromise the safety of the system. Due to timing uncertainties, i.e., incorrect clock synchronizations, there may exist various reasons why such real-time constraints are not met by a control system. 
Hence, a natural objective is to design a control system to be robust to such timing uncertainties and to even maximize this robustness. 
 To formulate  real-time constraints, researchers have used real-time temporal logics  as a  specification formalism such as Metric Temporal Logic (MTL) \cite{koymans1990specifying} and Signal Temporal Logic (STL) \cite{MalerN2004STL}. In this work, we focus on the  time robustness  of STL specifications defined in  \cite{Donze10STLRob} and propose and solve a control problem that aims to maximize the time robustness to make a system resilient to  timing uncertainties.

\subsection{Related Work}
\label{sec:related_work}

The real-time temporal logic STL, introduced in \cite{MalerN2004STL}, is interpreted over real-valued continuous-time signals. Various forms of robust semantics have been proposed for STL that quantify the extent by which an STL specification is satisfied or violated by a signal.  The authors in \cite{FainekosP09tcs} define the robustness degree as a  tube around a signal in which all signals either satisfy or violate the specification at hand. The robustness degree reflects a notion of space robustness. Various other space robustness notions have been proposed in the literature, such as \cite{Donze10STLRob} that closely follows  \cite{FainekosP09tcs}. Other space robustness notions are the arithmetic-geometric integral mean robustness~\cite{mehdipour2019average} and the smooth cumulative robustness \cite{haghighi2019control}. Another space robustness notion, especially tailored for aiding exploration in reinforcement learning problems, has been presented in \cite{varnai2020robustness}. A connection between linear, time-invariant filtering and space robustness  has been made in \cite{rodionova2016temporal}.  The authors in \cite{Donze10STLRob} make a first step towards defining a notion of time robustness that we particularly utilize in this paper. Alternatively, \cite{akazaki2015time} proposes averaged STL and captures a form of time robustness by averaging over time intervals and assigning priorities. This way, similarly to the time robustness in \cite{Donze10STLRob}, expeditiousness and other  modalities can be captured in terms of robustness. All these notions, except for time robustness in \cite{Donze10STLRob},  primarily focus on space robustness.


Control of dynamical systems under  STL specifications has first been considered by the authors in \cite{raman2014model} by proposing  a Mixed Integer Linear Program (MILP) encoding that allows to maximize the space robustness as defined in \cite{Donze10STLRob}. Other optimization-based methods, allowing again the maximization of space robustness, have been proposed in \cite{mehdipour2019average,pant2018fly,gilpin2020smooth} by using smooth approximations of the space robustness notion in  \cite{Donze10STLRob}  that allows to use of off-the-shelf gradient-based solvers. A reinforcement learning approach for multi-agent systems has been presented in \cite{muniraj2018enforcing}.  Another direction has been to develop robust feedback control laws that maximize the space robustness of an STL specification \cite{lindemann2018control}. While all these methods have presented compelling results, they focus on forms of space robustness while none of these methods maximize the time robustness of an STL specification. The closest work in terms of control under STL time robustness constraints is \cite{baras_runtime} where the authors, however, only consider very  special instances of STL (eventually and always operators). To the best of our knowledge, the problem of maximizing the time robustness by which a system satisfies an STL specification has neither been formulated nor solved. 

\subsection{Contributions and Paper Organization}
In this paper, we consider linear discrete-time systems and design controllers for which the time robustness of an STL specification, as presented in \cite{Donze10STLRob}, is maximized. In particular, we make the following contributions:

\begin{enumerate}
	\item We formulate a novel control problem to find an input sequence that maximizes the time robustness by which the system satisfies  an STL specification. Thereby, we define a new problem space for the  robust control of time-critical systems.
	\item We extend the theoretical foundation of time robustness and complement the work in  \cite{Donze10STLRob} with additional theoretical results on soundness of time robustness.
	\item To solve the optimal control problem, we propose a novel Mixed-Integer Linear Program (MILP) encoding. We provide correctness guarantees and a complexity analysis of the encoding. 
	\item We show how the proposed MILP encoding can be used to perform the time-robust control for unmanned aerial vehicles and multi-agent surveillance.
	
\end{enumerate}

The remainder of the paper is organized as follows. Sec.~\ref{sec:stl} introduces STL with its qualitative and time-robust semantics. 
	In Sec.~\ref{sec:prob_formulation} we state the control synthesis problem that we aim to solve. We also present the soundness theorem with the proof available in the Appendix Sec.~\ref{sec:proof_sat}. 
In Sec.~\ref{sec:milp} we present the proposed solution using the MILP encoding. Extensive simulations and case studies are presented in Sec.~\ref{sec:experiments}.
Finally, we summarize with conclusions in Sec.~\ref{sec:conclusions}.
\section{Signal Temporal Logic (STL) Robustness}
\label{sec:stl}

Let $\sstraj$ be a discrete-time signal $\sstraj:\TDom \to X$ such that $\TDom\subseteq\Ze_{\geq0}$ is the time domain and $X\subseteq\Re^n$ is a metric space. 
We denote a signal state at time step $t$ as $x_t\in X$ and call the set of all signals $\sstraj: \TDom \rightarrow X$ the \textit{signal space} $\SigSpace$.
Let $M=\{\mu_1,\ldots,\mu_L\}$ be a set of real-valued linear functions of the state $x$, $\mu_k(x):X \rightarrow \Re$.
For each $\mu_k$ its corresponding \textit{predicate} $p_k$ is defined as $p_k \defeq \mu_k(x) \geq 0$. 
Thus, each predicate defines a set in which $p_k$ holds true, namely $p_k$ defines the set $\{x\in X\such \mu_k(x)\geq 0\}$ in which $p_k$ is true.
All defined predicates construct the set $AP \defeq \{p_1,\ldots,p_L\}$.
Let interval $I=[a,b]\subset\TDom$ be a non-empty time interval where $0 \leq a \leq b$.
For $t\in\TDom$, the time interval $[t+a, t+b]$ is denoted as $t+I$.
The supremum operator is written $\sqcup$ and infimum is written $\sqcap$, $\top$ is a Boolean \textit{true}.
We interpret $\sign(0)=1$.


The syntax of Signal Temporal Logic (STL) is defined recursively as follows~\cite{MalerN2004STL}:
\begin{equation}
	\formula \defeq p\ |\ \neg \formula \ |\ \formula_1 \land \formula_2 \ |\ \formula_1 \until_I \formula_2
\end{equation}
where 
$p\in AP$ is a predicate, $\neg$ and $\land$ are the Boolean negation and conjunction, respectively, and $\until_I$ is the Until temporal operator over bounded interval $I$. 
The disjunction ($\lor$) and implication ($\implies$) are defined as usual. 
Additional temporal operators Eventually ($\eventually$) and Always ($\always$) can be defined as $\eventually_I\varphi = \top\until_I\varphi$ and $\always_I\varphi = \neg\eventually_I\neg\varphi$.

Formally, the \textit{semantics} of an STL formula $\formula$ defines what it means for a system trajectory $\sstraj$ to satisfy $\formula$ at time point $t$, denoted as $(\sstraj,t)\models \varphi$. 
If satisfaction does not hold it is denoted as $(\sstraj,t)\not\models \varphi$.
We will use the characteristic function notation:
\begin{definition}[STL characteristic function \cite{Donze10STLRob}]
	\label{def:char_func}
	The characteristic function $\chi_\varphi(\sstraj,t): \SigSpace \times \TDom \to \{-1,+1\}$ 
	of an  STL formula $\varphi$ relative to
	a trajectory $\sstraj$ at time $t$ is defined inductively as:
	\begin{equation}
		\label{eq:boolean_sat_char}
	\begin{aligned}
		\chi_p(\sstraj, t) &= 
		\sign(\mu(x_t))
		\\
		\chi_{\neg \formula}(\sstraj,t) &= -\chi_\formula(\sstraj,t)
		\\
		\chi_{\formula_1 \land \formula_2}(\sstraj,t) &= \chi_{\formula_1}(\sstraj,t) \ \sqcap\ \chi_{\formula_2}(\sstraj,t)
		\\
		\chi_{\formula_1 \until_I \formula_2}(\sstraj,t) &= 
		\bigsqcup_{t'\in t+I} \left(
		\chi_{\formula_2}(\sstraj,t') \ \sqcap \
		\bigsqcap_{t'' \in [t,t')} \chi_{\formula_1}(\sstraj,t'')
		\right)
	\end{aligned}
	\end{equation}
\end{definition}
%

The Boolean semantics of STL states that $\chi_\varphi(\sstraj,t)=1$ when $(\sstraj,t)\models\varphi$, and $\chi_\varphi(\sstraj,t)=-1$ when $(\sstraj,t)\not\models \varphi$.  
While the Boolean STL semantics shows \textit{whether} a signal
$\sstraj$
satisfies a given specification $\varphi$ at time $t$
or
not, STL quantitative semantics, also known
as robustness, measures \textit{how much} the signal is satisfying or
violating the specification.
First notion of a quantitative measure of satisfaction has been presented in \cite{FainekosP09tcs}. This notion defines a so-called \textit{spatial robustness degree}.
Later an alternative notion was presented in \cite{Donze10STLRob} where the authors suggested the novel measure of \textit{time robustness}.

\begin{definition}[Time Robustness \cite{Donze10STLRob}]
	\label{def:time_rob_rec}
	The right and left time robustness of an STL formula $\varphi$ with respect to a trajectory $\sstraj$ at time $t$ are defined inductively as:
	\begin{equation}
		\label{eq:thetap}
		\begin{aligned}
			\thetap_p(\sstraj,t) = \chi_p(\sstraj, t)\cdot
			\max\{\tau\geq0\ :\ &\forall t'\in[t,t+\tau],\\ &\chi_p(\sstraj,t')=\chi_p(\sstraj,t)\} 
		\end{aligned}
	\end{equation}
	\begin{equation}
		\label{eq:thetam}
		\begin{aligned}
			\thetam_p(\sstraj,t) = \chi_p(\sstraj, t)\cdot
			\max\{\tau\geq0\ :\ &\forall t'\in[t-\tau,t],\\ &\chi_p(\sstraj,t')=\chi_p(\sstraj,t)\} 
		\end{aligned}
	\end{equation}	
	and then applying to each $\theta^{\bowtie}_p$, where $\bowtie\ \in\{+,-\}$, the recursive rules of the operators similarly to Def.~\ref{def:char_func} resulting in:
	\begin{align}
		\label{eq:t_neg}
		\theta^{\bowtie}_{\neg \formula}(\sstraj,t) &= -\theta^{\bowtie}_\formula(\sstraj,t)
		\\
		\label{eq:t_and}
		\theta^{\bowtie}_{\formula_1 \land \formula_2}(\sstraj,t) &= \theta^{\bowtie}_{\formula_1}(\sstraj,t) \ \sqcap\ \theta^{\bowtie}_{\formula_2}(\sstraj,t)
		\\
		\label{eq:t_until}
		\theta^{\bowtie}_{\formula_1 \until_I \formula_2}(\sstraj,t) &= 
		\bigsqcup_{t'\in t+I} \left(
		\theta^{\bowtie}_{\formula_2}(\sstraj,t') \ \sqcap \
		\bigsqcap_{t'' \in [t,t')} \theta^{\bowtie}_{\formula_1}(\sstraj,t'')
		\right)
	\end{align}
\end{definition}

\begin{figure}[t!]
	\centering
	\includegraphics[width=.5\textwidth]{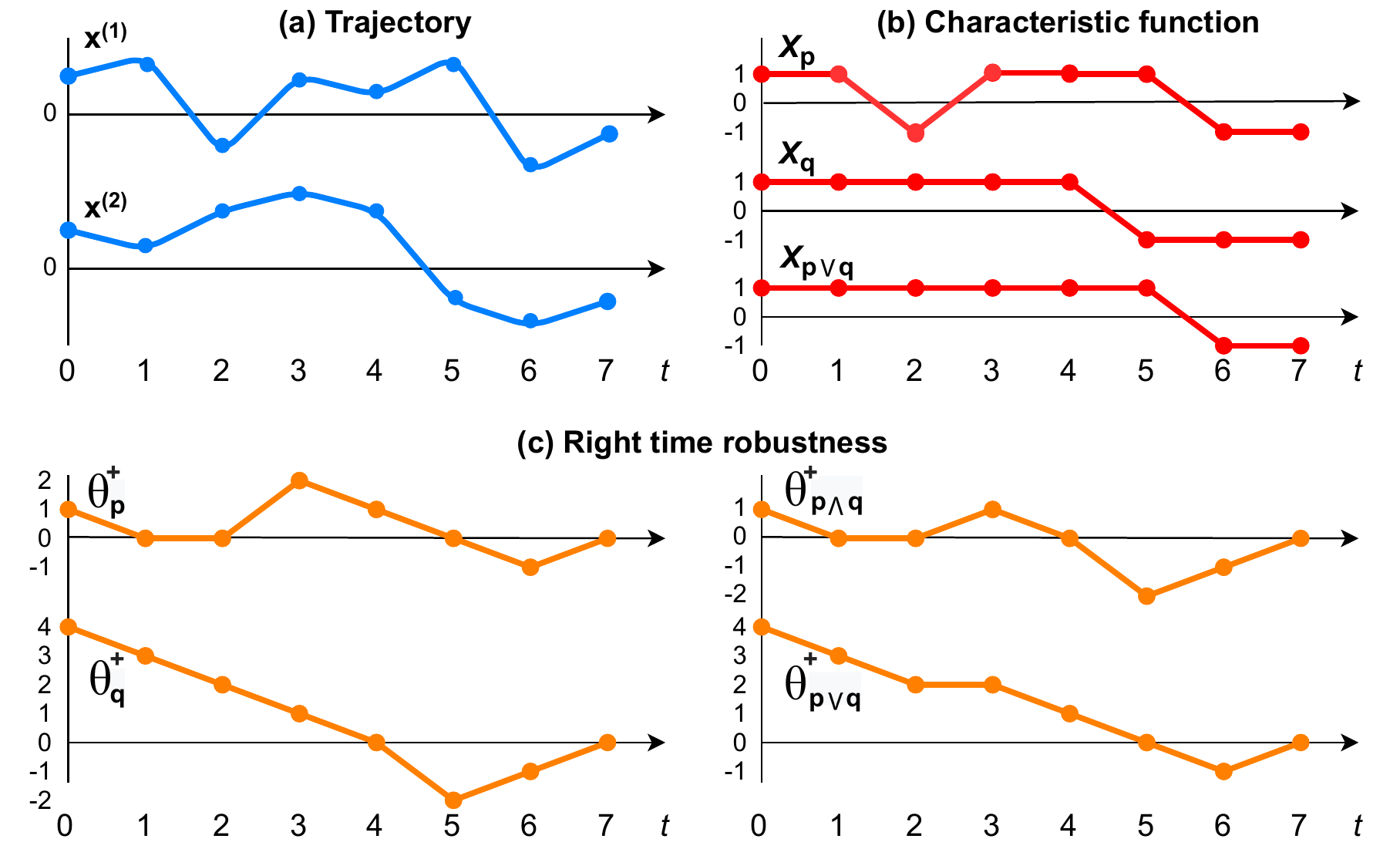}
	\caption{Evolution of the trajectory (a), characteristic function (b) and time robustness (c) from Example \ref{ex:running}.}
	\label{fig:ex1}
	\vspace{-5mm}
\end{figure}


\begin{exmp}
	\label{ex:running}
	We next illustrate the time robustness notion in  Fig.~\ref{fig:ex1}. 
	The trajectory $\sstraj=[\sstraj^{(1)},\, \sstraj^{(2)}]^\top$ presented in Fig.\ref{fig:ex1}(a) is finite and discrete-time, its state at each time step $t=0,\ldots,7$ is $x_t=[x_t^{(1)},\,x_t^{(2)}]^\top\in\Re^2$.
	Let the two predicates be $p=x^{(1)}\geq 0$ and $q=x^{(2)}\geq 0$. The characteristic functions for $p$, $q$ and $\varphi=p\vee q$ 
	are shown in Fig.~\ref{fig:ex1}(b). 
	The evolutions of the right time robustness $\thetap_\varphi(\sstraj,t)$, $t=0,\ldots,7$ for the above predicates, $\varphi=p \vee q$ and $\varphi=p\wedge q$ are presented in Fig.~\ref{fig:ex1}(c). For instance, consider time step $t=3$. Since $\chi_p(\sstraj,3)=\chi_p(\sstraj,4)=\chi_p(\sstraj,5)=+1$ and $\chi_p(\sstraj,6)=-1$, by Def.~\ref{def:time_rob_rec}, $\thetap_{p}(\sstraj,3)=2$. On the other hand, $\chi_q(\sstraj,3)=\chi_q(\sstraj,4)=+1\not = -1 = \chi_q(\sstraj,5)$, thus, $\thetap_{q}(\sstraj,3)=1$. 
	Also, from \eqref{eq:t_and}, 
	$\thetap_{p\wedge q}(\sstraj,3) = \min (\thetap_{p}(\sstraj,3),\, \thetap_{q}(\sstraj,3)) = 1$ but
	$\thetap_{p\vee q}(\sstraj,3) = \max (\thetap_{p}(\sstraj,3), \thetap_{q}(\sstraj,3)) = 2$.
\end{exmp}
\subsection{Soundness of Time Robustness}

In this section, we formulate and prove the theorem that 
states the relationship between the time robustness and Boolean semantics of STL. 
The following Theorem~\ref{thm:sat} is fundamental to the underlying theory of time robustness and the control synthesis problem defined in the next section, but to the best of our knowledge has never been formally stated in any previously published papers. 

 \begin{theorem}[Soundness]
	\label{thm:sat}
	For an STL formula $\varphi$, trajectory $\sstraj:\TDom \rightarrow X$, time $t\in\TDom$ and $\bowtie\in\{+,-\}$ the following results hold: 
	\begin{enumerate}
		\item\label{i1} $\theta^{\bowtie}_\varphi(\sstraj,t) > 0 \quad\, \Longrightarrow\ \chi_\varphi(\sstraj,t)= +1$
		\item\label{i2} $\theta^{\bowtie}_\varphi(\sstraj,t) < 0 \quad\, \Longrightarrow\ \chi_\varphi(\sstraj,t)= -1$
		\item\label{i3} $\chi_\varphi(\sstraj,t)= +1 \  \Longrightarrow\ 
		\theta^{\bowtie}_\varphi(\sstraj,t) \geq 0$
		\item\label{i4} $\chi_\varphi(\sstraj,t)= -1 \  \Longrightarrow\ 
		\theta^{\bowtie}_\varphi(\sstraj,t) \leq 0$ 
	\end{enumerate}
\end{theorem}
See the proof in the Appendix Sec.~\ref{sec:proof_sat}.

Note that equivalence $\theta^{\bowtie}_\varphi(\sstraj,t) \geq 0 \Longleftrightarrow \chi_\varphi(\sstraj,t)= +1$ does not hold\footnote{Equivalence $\theta^{\bowtie}_\varphi(\sstraj,t) \leq 0 \Longleftrightarrow \chi_\varphi(\sstraj,t)= -1$ does not hold either, same reasoning applied.} since when $\theta^{\bowtie}_\varphi(\sstraj,t) =0$ we cannot determine if the formula is satisfied or violated. 

\begin{exmp}[continues=ex:running]
	Take a look again at the trajectory $\sstraj$ shown in Fig.~\ref{fig:ex1}.
	Consider $\varphi= p \vee q$.
	From Fig~\ref{fig:ex1}(c) one can see that 
	$\thetap_{\varphi}(\sstraj,3)= 2 > 0$ and
	$\thetap_{\varphi}(\sstraj,6)= -1 < 0$.
	Since 
	$\chi_{\varphi}(\sstraj,3) = +1$ then due to Thm.~\ref{thm:sat}, $(\sstraj,3)\models \varphi$. Similarly, $\chi_{\varphi}(\sstraj,6) = -1$ leads to $(\sstraj,6)\not\models \varphi$.
	By taking a look at Fig.~\ref{fig:ex1}(b) one can conclude, that these facts indeed hold.	
	Also note that for time points $t=5$ and $t=7$, 
	$\thetap_\varphi(\sstraj,5)=\thetap_\varphi(\sstraj,7)=0$ but $\chi_\varphi(\sstraj,5)= +1$ and  $\chi_\varphi(\sstraj,7)= -1$.
\end{exmp}

\section{Time-Robust STL Control Synthesis}
\label{sec:prob_formulation}


Consider a discrete-time, linear control system: 
\begin{equation}
	\label{eq:model}
	x_{t+1} = A x_t + B u_t
\end{equation}
where $x_t \in X \subseteq \Re^n$ is the current state of the system in a bounded domain $X$,
$u_t \in U \subseteq \Re^m$ is the current control input, $A\in\Re^{n\times n}$ and $B\in\Re^{n\times m}$.
The system's initial state $x_0$ takes values from some initial set $X_0 \subseteq X$.
Given an initial state $x_0$ and a finite control input sequence $\inpSig = (u_0,u_1\ldots, u_{H-1})$ s.t. $u_{t} \in U$, 
a \textit{trajectory} of the system is the unique sequence of states $\sstraj = (x_0,x_{1}\ldots,x_{H})$, s.t.
$x_{t}\in X$ and \eqref{eq:model} holds.
We denote $\TDom = \{0, 1,\ldots, H\}$ to be a finite discrete time domain, where $H$ is a time horizon 
which is sufficiently large to verify the satisfaction of formula $\varphi$.


For time-critical systems one is often not only interested in satisfying an STL specification $\varphi$ but also in satisfying $\varphi$ robustly with respect to the time robustness  $\theta^{\bowtie}_{\varphi}(\sstraj,0)$. Achieving such robustness is  particularly important when the system is subject to timing uncertainties, including agent delays or early starts. Towards this goal, we aim to maximize time robustness $\theta^{\bowtie}_{\varphi}(\sstraj,0)$ while imposing a lower bound $\theta^*$ on $\theta^{\bowtie}_{\varphi}(\sstraj,0)$. In other words, for a given STL specification $\varphi$ and initial condition $x_0\in X_0$, we want to find a
control input sequence $\inpSig^* = (u^*_0,u^*_1\ldots, u^*_{H-1})$ such that the corresponding system trajectory $\sstraj^*$ 
satisfies the specification $\varphi$ and results in a time robustness $\theta^{\bowtie}_\varphi(\sstraj^*,0)$ that is maximized and satisfies a minimum required time robustness $\theta^*>0$. 
Formally, this can be defined as the following 
problem. 

\begin{prob}[Time-Robust STL Control Synthesis]
	\label{prob:synthesis}
Given an STL specification $\varphi$, time horizon $H$, discrete-time linear control system \eqref{eq:model} with initial condition $x_0\in X_0$ and a lower bound $\theta^*$, solve
	\begin{equation*}
		\label{eq:control_problem}
		\begin{aligned}
			\inpSig^*=
			\underset{\inpSig}{\argmax}
			&\quad \theta^{\bowtie}_\varphi(\sstraj,0) \\
			\text{s.t.}
			&\quad x_{t+1} = A x_t + B u_t,\ u_t\in U,\ t=0,\ldots,H-1\\
			&\quad x_t\in X,\ t=0,\ldots,H\\
			&\quad \theta^{\bowtie}_\varphi(\sstraj,0)\geq \theta^*>0.
		\end{aligned}
	\end{equation*} 
\end{prob}


Since the robustness function $\theta^{\bowtie}_\varphi$ is neither continuous nor smooth,  gradient-based solvers cannot be applied to solve Prob.~\ref{prob:synthesis}.
Even more challenging is the fact that $\theta^{\bowtie}_\varphi$ includes signal shifts according to \eqref{eq:thetap}-\eqref{eq:thetam}. Thus, techniques based on smooth approximations \cite{pant2017smooth,pant2018fly}, non-smooth optimization theory \cite{AbbasF_NonlinearDescent13} or Monte-Carlo optimization  \cite{AbbasFSIG13tecs} are not applicable either. This motivates the use of Mixed-Integer Linear Programming (MILP) in this work  to explicitly encode the signal shifts in \eqref{eq:thetap}-\eqref{eq:thetam}. We describe the details of the MILP encoding of Prob.~\ref{prob:synthesis} in the next section. 

\section{MILP ENCODING OF TIME ROBUSTNESS}
\label{sec:milp}

In this section, we present the right time robustness encoding. The left time robustness can be encoded analogously with only minor modifications and is hence omitted. 

Following Def.~\ref{def:time_rob_rec}, given the STL formula $\varphi$, the time robustness can be computed recursively on the structure of $\varphi$. 
We will start with the main milestone of the overall time robustness MILP encoding, that is the encoding of predicates, i.e. $\thetap_{p}(\sstraj,t)$, in Section \ref{sec:milp_pred}. We  then briefly describe the encoding of other STL operators in Section \ref{sec:milp_op}.

\begin{algorithm}[t!]
	\SetAlgoLined
	\KwIn{Linear predicate $p := \mu(x)\geq 0$, trajectory $\sstraj$}
	\KwOut{Right time robustness $\thetap_{p}(\sstraj)$ and the set of MILP constraints $\mathcal{P}$}

	{1: Let $z_t\in\Be$ be constrained by \eqref{eq:mu}, $t=0,\ldots,H$.}
	
	{2: Let $\chi_p(\sstraj)$ be constrained by \eqref{eq:chi_}.}
	
	{3: Let $c_t^1, c_t^0\in\Ze$  be constrained by \eqref{eq:c1} and \eqref{eq:c0}.}
	
	{4: $\thetap_{p}(\sstraj)\overset{\eqref{eq:milp_thetap}}{=}c_t^1 + c_t^0 - \chi_p(\sstraj,t)$.}
	
	{5: $\mathcal{P}$ consists of the MILP constraints \eqref{eq:mu}, \eqref{eq:chi_},  \eqref{eq:c1}, \eqref{eq:c0}, and \eqref{eq:milp_thetap}.}
	
	\caption{The function $(\thetap_{p}(\sstraj),\mathcal{P})=\texttt{MILP\_PREDICATE}(p,\sstraj)$}
	\label{alg:pred}
\end{algorithm}

\subsection{MILP Encoding of STL Predicates} 
\label{sec:milp_pred}

We propose the idea of using  counter variables within an MILP that enumerate the sequence of interest. Such counting idea implemented through the proposed MILP counters described later by \eqref{eq:c1}-\eqref{eq:c0} is the key concept behind the presented time robustness MILP encoding.
Below we first summarize the result of the encoding for STL predicates, then present an algorithm where we explain the construction in more details and finally, consider the encoding on a particular example.

\begin{prop}[MILP encoding of STL predicates]
	For a linear predicate $p\in AP$ and trajectory $\sstraj$ described by the system constraints \eqref{eq:model}, the right time robustness sequence $\thetap_p\!(\sstraj)\!\!=\!\!(\thetap_p\!(\sstraj,0),\!\ldots\!,\!\thetap_p\!(\sstraj,H))$, with $\thetap_p\!(\sstraj,t)$ defined by \eqref{eq:thetap}, is equivalent to the set of MILP constraints produced by the function $(\thetap_{p}(\sstraj),\mathcal{P})=\texttt{MILP\_PREDICATE}(p,\sstraj)$ in Alg.~\ref{alg:pred}.
\end{prop}

The proof is by construction and we explain a step-by-step construction of the Alg.~\ref{alg:pred} as follows. 

\begin{enumerate}
	\item First, we construct a binary variable $z_t\in \{0,1\}$ that corresponds to the Boolean satisfaction of the predicate $p$ by trajectory $\sstraj$ at every time point $t=0,\ldots,H$ , i.e. we enforce that $z_t=1$ if and only if $\mu(x_t)\geq 0$. With an assumption that $\mu(x_t)$ is a linear function of the state, variable $z_t$ can be defined as a set of MILP constraints as follows~\cite{bemporad1999control}:
	\vspace{-4pt}
	\begin{equation}
		\label{eq:mu}
		\mu(x_t) \leq (M+\epsilon)\cdot z_t - \epsilon, \quad
		\mu(x_t) \geq m\cdot (1-z_t),
		\vspace{-4pt}
	\end{equation}
	where $\epsilon$ is a small positive constant that represents the tolerance, $M = \max_{x\in X} \mu(x)$ and $m=\min_{x\in X} \mu(x)$. By \cite{bemporad1999control}, the over-estimate of $M$ and under-estimate of $m$ suffice for the equivalence as well. 
	
	\item Since $z_t=1$ if and only if $\mu(x_t)\geq 0$ and by Def.~\ref{def:char_func} $\chi_p(\sstraj,t)=\sign \mu(x_t)\in\{\pm1\}$, the characteristic function can be encoded as 
	\vspace{-4pt}
	\begin{equation}\label{eq:chi_}
	\chi_{p}(\sstraj,t) = 2 z_t -1.
	\vspace{-4pt}	
	\end{equation}
	
	\item\label{it:l}  Recall that
	\begin{align*}
		\thetap_p(\sstraj,t) = \chi_p(\sstraj, t)\cdot
		\max\{\tau\geq0\ :\ &\forall t'\in[t,t+\tau],\\ &\chi_p(\sstraj,t')=\chi_p(\sstraj,t)\}. 
	\end{align*}
	Since $\chi_p(\sstraj,t) \in\{-1,+1\}$ there exist two disjoint possibilities for $\thetap_p$ which in terms of $z_t$ can be written as:
	\begin{equation*}
		\thetap_p(\sstraj,t)\! =\!\!  
		\begin{cases}
			\max\{\tau:\forall t'\in[t,t+\tau], z_{t'}=1\} ,&\text{if } z_t=1 \\
			-\max\{\tau:\forall t'\in[t,t+\tau], z_{t'}=0\},&\text{if } z_t=0 \\
		\end{cases}
	\end{equation*}
where $\tau\geq 0$.
	In other words, if at time point $t$, $z_t=1$, one can count the maximum number of sequential $z_{t'}=1$ where $t'>t$ in order to calculate $\thetap_p(\sstraj,t)$. On the other hand, if $z_t=0$, then counting 0s to the right and multiplying the final value by $-1$ will define the time robustness value. 
	
	
	\item\label{it:c} To implement the counting idea mentioned in the previous step, we construct two variables that count 1s and 0s to the right of $t$ but for $t'\geq t$.
	Let their recursive definitions be as following: 
		\begin{align}
		\label{eq:c1}
		c_t^1 &= (c_{t+1}^1+1) \cdot z_t, \qquad\qquad\ \, c^1_{H+1}=0\\
		\label{eq:c0}
		c_t^0 &= (c_{t+1}^0-1) \cdot (1-z_t),
		\qquad c^0_{H+1}=0
	\end{align} 
	By construction, the counter $c_t^1$ counts the maximum number of sequential $z_{t'}=1$ when $t'\geq t$.
	The second counter $c_t^0$ counts sequential $z_{t'}=0$ when $t'\geq t$ and multiplies the final value by $-1$. 
	Note that counters are defined backwards: from $t=H+1$ to $t=0$, so 
	while the length of trajectory $\sstraj$ is $H+1$,
	the lengths of $c_t$ sequences are $H+2$. 
	
	\item 
	Time robustness is defined by the sequential $z_{t'}$ when $t'>t$, not when  $t'\geq t$. 
	Therefore, to use the counters \eqref{eq:c1}-\eqref{eq:c0} to encode the time robustness, the counters should be modified as they must exclude the corresponding counted value at time point $t$:
	\begin{align*}
		\tilde{c}_t^1 &= c_t^1 - z_t, \qquad\quad\ \, t=0,\ldots,H\\
		\tilde{c}_t^0 &= c_t^0 + (1-z_t),\quad t=0,\ldots,H
	\end{align*}
	
	\item Using the fact from Step \ref{it:l}) that two possibilities of $z_t$ being 1 or 0 are disjoint (i.e. $\tilde{c}_t^1 \neq 0$ then $\tilde{c}_t^0 = 0$), the right time robustness $\thetap_p(\sstraj,t)$ is defined as:
	\begin{align}
		\thetap_p(\sstraj,t) &= \tilde{c}_t^1 + \tilde{c}_t^0 = c_t^1 + c_t^0 - (2z_t -1) \nonumber\\
		& = c_t^1 + c_t^0 - \chi_p(\sstraj,t). \label{eq:milp_thetap}
	\end{align} 
\end{enumerate}

\begin{table}[t!]
	\renewcommand{\arraystretch}{1.5}
	\setlength{\tabcolsep}{3pt}
	\centering
	\begin{tabular}{|l|c|c|c|c|c|c|c|c|c|}
		\hline
		$t$   & 0 & 1  & 2  & 3 & 4 & 5 & 6 & 7 & 8 \\ \hline
		\hline
		$z_t$ & 1 & 1 & 0 & 1 & 1 & 1 & 0 & 0 &  \\ \hline
		$\chi_p(\sstraj,t)$ & 1 & 1 & -1 & 1 & 1 & 1 & -1 & -1 &  \\ \hline
		$c^1_t =(c_{t+1}^1+1) \cdot z_t,\  c^1_{H+1}=0$    & 2 & 1  &  0 & 3  &  2 &  1 &  0 & 0  & 0  \\ \hline
		$c_t^0 = (c_{t+1}^0-1) \cdot (1-z_t),
		\ c^0_{H+1}=0$    &  0  &  0  & -1  & 0  & 0  & 0  & -2  &  -1 & 0  \\ \hline\hline
		$\thetap_p(t) = c^1_t + c^0_t - \chi_p(\sstraj,t)$ & 1   & 0  & 0  & 2  & 1  &  0 & -1  & 0 &   \\ \hline
	\end{tabular}
	\vspace{0pt}
	\caption{\small Estimation of $\thetap_p\!(\sstraj,t)$ from Example~\ref{ex:running} following Alg.~\ref{alg:pred}.}
	\label{tab:exmp_theta}
	\vspace{-6mm}
\end{table}

\begin{exmp}[continues=ex:running]
	Consider again the signal $\sstraj$ and predicate $p$ shown in Fig.~\ref{fig:ex1}.
	In Table~\ref{tab:exmp_theta} we consider a step-by-step estimation of the right time robustness for predicate $p$ following the MILP encoding procedure described in Alg.\ref{alg:pred} and Sec.\ref{sec:milp_pred}.
	On the other hand, using Def.~\ref{def:time_rob_rec}, one can check that $\thetap_p(\sstraj)$ is indeed equal to $(1, 0, 0, 2, 1, 0, -1, 0)$. MILP encoding procedure leads to the same result as its estimation by the definition.
\end{exmp}

\textit{Remark.} Constraints \eqref{eq:c1}-\eqref{eq:c0} are specified using a product of integer and Boolean variables. In Lemma~\ref{lemma:xz} below we show that such product can be expressed as MILP constraints.
\begin{lemma}[{If-then-else product construct~\cite{bemporad2001discrete}}]
	\label{lemma:xz}
	Let $b\in\{0,1\}$ be a Boolean variable and let $x$ be an integer variable such that lower and upper bounds are known constants, $x_l\leq x\leq x_{u}$. The expression 
	$y=b\cdot x$ can be equivalently expressed as a set of mixed-integer linear constraints as follows:
	\begin{equation}
		\label{eq:inq}
		\begin{aligned}
			x_lb\leq\  &y\leq x_{u} b \\
			x - x_{u}(1-b) \leq &y \leq x- x_l(1-b).
		\end{aligned}
	\vspace{-4pt}
	\end{equation}
\end{lemma}

\begin{table*}[t!]
	\renewcommand{\arraystretch}{1.2}
	\setlength{\tabcolsep}{5.pt}
	\centering
	\begin{tabular}{l|c|cc|cc|c}
		\multicolumn{1}{c|}{\multirow{2}{*}{\textbf{Mission}}} &
		\multicolumn{1}{c|}{\multirow{2}{*}{\textbf{\# Constraints}}} &
		\multicolumn{2}{c|}{\textbf{\# Variables}} &
		\multicolumn{2}{c|}{\textbf{Computation time (s)}} &
		\multicolumn{1}{c}{\textbf{Time Rob., $\thetap_{\formula}(\sstraj,0)$}} \\
		\multicolumn{1}{c|}{} &
		\multicolumn{1}{c|}{} &
		\multicolumn{1}{c}{\textbf{Boolean}} &
		\multicolumn{1}{c|}{\textbf{Integer}} &
		\multicolumn{1}{c}{\textbf{YALMIP}} &
		\multicolumn{1}{c|}{\textbf{Solver}} &
		\multicolumn{1}{c}{(Time units)} \\ \hline
		$\varphi_1 = \always_{[0,20]} p$, Sec.~\ref{sec:experiments} & 543 & 71 & 151 & 0.26 &0.15 & 29  \\
		$\varphi_2 = \eventually_{[0,20]} q$, Sec.~\ref{sec:experiments} & 543 & 71 & 119 & 0.25 & 0.18 & 49 \\
		$\varphi_3 = \always_{[0,10]}\eventually_{[0,10]} p$, Sec.~\ref{sec:experiments} & 776 & 182 & 162 & 0.27 & 0.28 & 39 \\
		$\varphi_4 = \varphi_2\wedge \always_{[0,5]} p$, Sec.~\ref{sec:experiments} & 1061 & 129 & 253 & 0.28 & 0.23 & 44 \\
		$\varphi_5 = \varphi_2 \wedge
		\varphi_3$, Sec.~\ref{sec:experiments} & 1324 & 255 & 264 & 0.31  & 0.24 & 39 \\
		\hline
		Case study 1, $\varphi_{uav}$,  Eq.\eqref{eq:ex_drone} &2447  &224  &502  &0.41  & 0.34&23 \\ \hline
		Case study 2, $\varphi_{surv}$, Eq.\eqref{eq:mission2}& 8794 & 2750 & 858 & 0.82 & 49.69&14 \\
		Case study 2, $\varphi'_{surv}$, Eq.\eqref{eq:vel_constr} & 9034 & 2750 & 858 & 0.88 &29.08 & 14 \\
		Case study 2, feasibility formulation, $\theta_{\varphi_{surv}}(\sstraj,0)=5$, Fig.~\ref{fig:ex_agents(t)}(a) & 8795 & 2750 &858  & 0.81 & 2.12 & 5\\
		Case study 2, feasibility formulation, $\theta_{\varphi_{surv}}(\sstraj,0)=1$, Fig.~\ref{fig:ex_agents(t)}(b) & 8795 & 2750 &858  & 0.85 & 1.02 & 1
	\end{tabular}
	\caption{\small Computational complexity report. \textit{Computation time} includes \textit{YALMIP time} (the time used to build the MILP and convert it into appropriate format for the solver) and \textit{Solver time} (the time taken by Gurobi to solve the problem).}
	\label{tab:experiments}
	\vspace{-6mm}%
\end{table*}

\subsection{MILP Encoding of STL Operators}
\label{sec:milp_op}

Having encoded the time robustness of STL predicates as MILP constraints, the generalization to STL formulas is straight-forward and can use the encoding from \cite{raman2014model}. 

For instance, let $\varphi = \wedge_{i=1}^n \varphi_i$ and $\forall i,\ \thetap_{\varphi_i}(\sstraj,t) = r_i$. Then $\thetap_{\formula}(\sstraj,t) = r$ if and only if:
\begin{equation}
	\label{eq:conj}
	\begin{aligned}
		&r_i - M(1-b_i) \leq r \leq r_i, \ \forall i\in\{1,\ldots n\}\\
		& \sum_{i=1}^{n} b_i = 1
	\end{aligned}
\end{equation}
where $b_i=\{0,1\}$, $i=\{1,\ldots,n\}$ are introduced binary variables and $M$ is a big-$M$ parameter.

For the complete definition of other STL operators, see the quantitative encoding of STL constraints from \cite{raman2014model}.

The overall MILP encoding framework is summarized in Alg.~\ref{alg:milp}. The function $\texttt{SYSTEM\_CONSTRAINTS}(x_0,\inpSig)$ defines linear constraints on the decision variable $u_t$ according to \eqref{eq:model} and such that $x_t\in X$. The function $\texttt{MILP\_PREDICATE}(p_k,\sstraj)$ is defined by Alg.~\ref{alg:pred}. 
	We denote all encoded predicates used in $\varphi$ for all time steps as $\thetap_{p}(\sstraj)$.
	The function $\texttt{MILP\_OPERATORS}(\varphi,\sstraj,\thetap_{p}(\sstraj))$ is defined according to \cite[Section IV.D.]{raman2014model} and recursively follows the structure of $\varphi$.
	It outputs $\thetap_{\varphi}(\sstraj,0)\in\Ze$ and a set of MILP constraints $\mathcal{O}$. For example, for conjunctions, $\mathcal{O}$ is illustrated in \eqref{eq:conj}.

We summarize the main properties of the above MILP formulation in the following proposition.
\begin{prop}
	\label{prop:complexity}
	\begin{enumerate}
		\item The time-robust STL control synthesis Problem~\ref{prob:synthesis} is equivalent to a MILP  and can be solved as described in Alg.~\ref{alg:milp}.
		\item MILP encoding of $\theta_\varphi^{\bowtie}(\sstraj,t)$ is a function of $O(H\cdot |AP|+|\varphi|)$ binary and continuous variables, where $H$ is the time horizon, $|\varphi|$ is the number of operators in the formula $\varphi$ and $|AP|$ is the number of used predicates.  
	\end{enumerate}
\end{prop}
%

\begin{algorithm}[b!]
	\SetAlgoLined
	\KwIn{Specification $\varphi$, initial state $x_0$, time horizon $H$, minimum time robustness $\theta^*$}
	\KwOut{Control input sequence $\inpSig^* = (u^*_0,\ldots, u^*_{H-1})$}
	
	Set $\inpSig=(u_0,\ldots,u_{H-1})$ to be the decision variable
	
	$\sstraj=\texttt{SYSTEM\_CONSTRAINTS}(x_0,\inpSig)$
	
	\For{$k=1,\ldots,L$}{
		
		$(\thetap_{p_k}(\sstraj),\mathcal{P}_k)=\texttt{MILP\_PREDICATE}(p_k,\sstraj)$ // Sec.~\ref{sec:milp_pred}
	}
	
	$(\thetap_{\varphi}(\sstraj,0),\mathcal{O}) =\texttt{MILP\_OPERATORS}(\varphi,\sstraj,\thetap_{p}(\sstraj))$   // Sec.~\ref{sec:milp_op}

	$\,\inpSig^*=
		\underset{\inpSig}{\argmax}\quad \theta^{+}_\varphi(\sstraj,0)$
	 
	 $\qquad\quad\ 
	 \text{s.t.}\quad\ \ \,\theta^{+}_\varphi(\sstraj,0)\geq \theta^*>0$
	 
	 $\qquad\qquad\qquad\ \,\sstraj=\texttt{SYSTEM\_CONSTRAINTS}(x_0,\inpSig)$
	 
	 $\qquad\qquad\qquad\ \,u_t\in U$ for $t=0,\hdots,H-1$
	 
	 $\qquad\qquad\qquad\ \,\mathcal{O},\ \mathcal{P}_k$, for $k=1,\hdots,L$
	
	\caption{MILP encoding of time robustness}
	\label{alg:milp}
\end{algorithm}



\section{EXPERIMENTAL RESULTS}
\label{sec:experiments}

The complete MILP implementation described in this paper together with the code to reproduce all the case studies presented in this section can be found as an open-source project under the following URL:  \href{https://github.com/nellro/time-robust-control}{https://github.com/nellro/time-robust-control}.


All simulations were performed on a computer with an Intel Core i7-9750H 6-core processor
and 16GB RAM, running Ubuntu 18.04. 
The MILP was implemented in MATLAB using YALMIP~\cite{lofberg2004yalmip} with Gurobi 9.1 solver~\cite{gurobi}.

Table~\ref{tab:experiments} presents the report on the computational complexity of all simulations described below. It includes the number of generated MILP constraints, number of created binary and integer variables and computation times that include YALMIP and solver times.

To analyze and compare the computational complexity of the two case studies presented below we first provide the results for the following five generic STL formulas:
\begin{gather*}
	\varphi_1 = \always_{[0,20]} p,\quad
	\varphi_2 = \eventually_{[0,20]} q,\\
	\varphi_3 = \always_{[0,10]}\eventually_{[0,10]} p,\quad
	\varphi_4 = \varphi_2\wedge \always_{[0,5]} p,\quad 
	\varphi_5 = \varphi_2 \wedge
	\varphi_3	
\end{gather*}
where $p = x \geq 0.1$, $q = x \leq 0.5$. We used a simple discrete-time linear system $x_{t+1}=u_t$, where $x_t,u_t\in\Re$, and set time horizon to $H=50$.
Table~\ref{tab:experiments} shows that with an increasing complexity of the formula, increases the number of constraints together with the number of created variables. This result is expected due to the previously stated Prop.~\ref{prop:complexity}. On the other hand, since solvers use various sophisticated
heuristics, solver time might not be directly connected to the complexity of the formula. For example, $\varphi_5 = \varphi_2 \wedge \varphi_3$ but the solver time for $\varphi_5$ is lower than for its both sub-formulas. 

\begin{figure*}[t!]
	\centering
	\begin{subfigure}[t]{1\columnwidth}
		\centering
		\includegraphics[width=\textwidth]{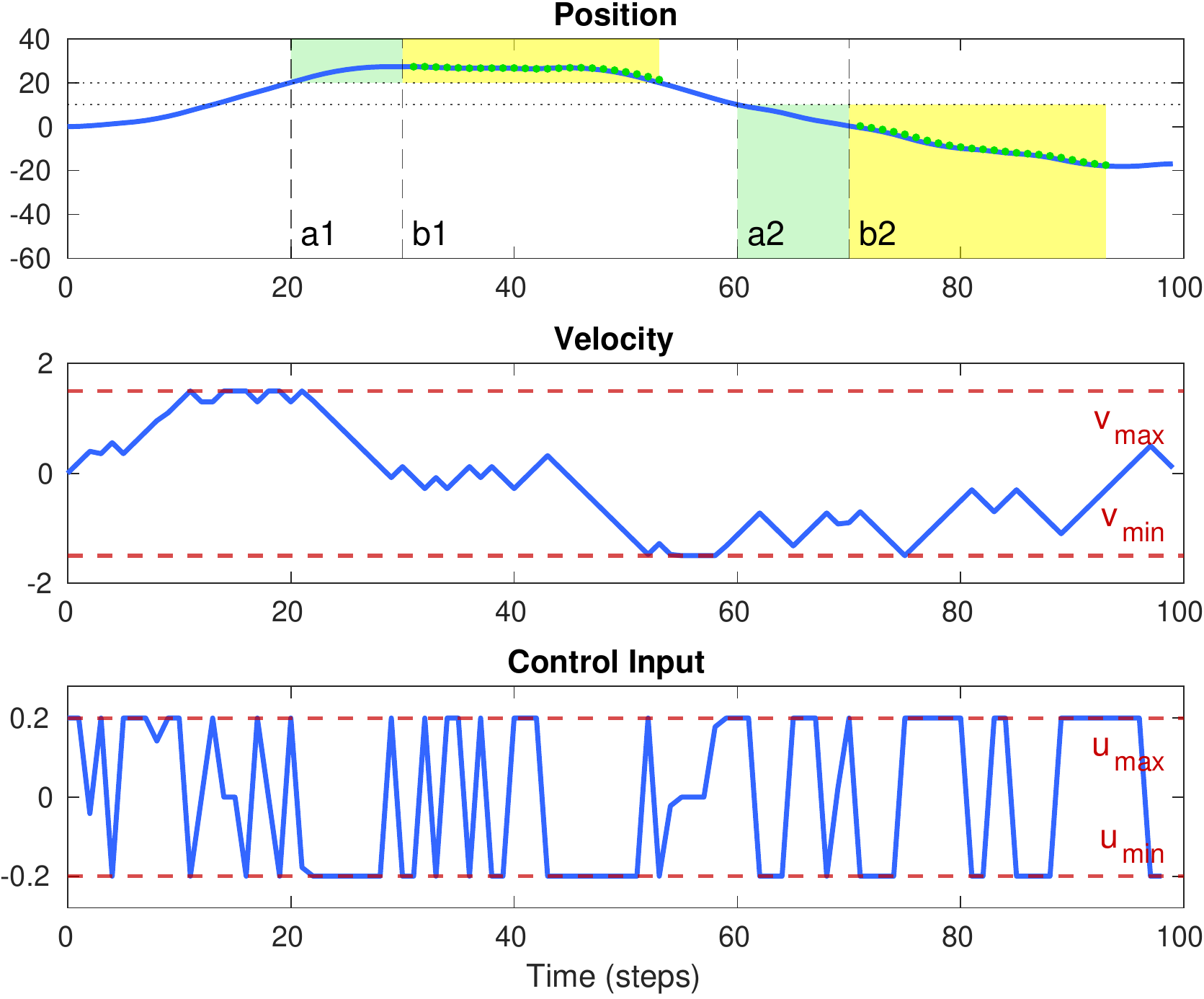}
		\caption{UAV position, velocity and control input. Both times, after staying within the green zone from $a_i$ to $b_i$ time steps, UAV continued staying there for 23 more steps (shown as green dots).}
	\end{subfigure}
	\hspace{5pt}
	\begin{subfigure}[t]{1\columnwidth}
		\centering
		\includegraphics[width=\textwidth]{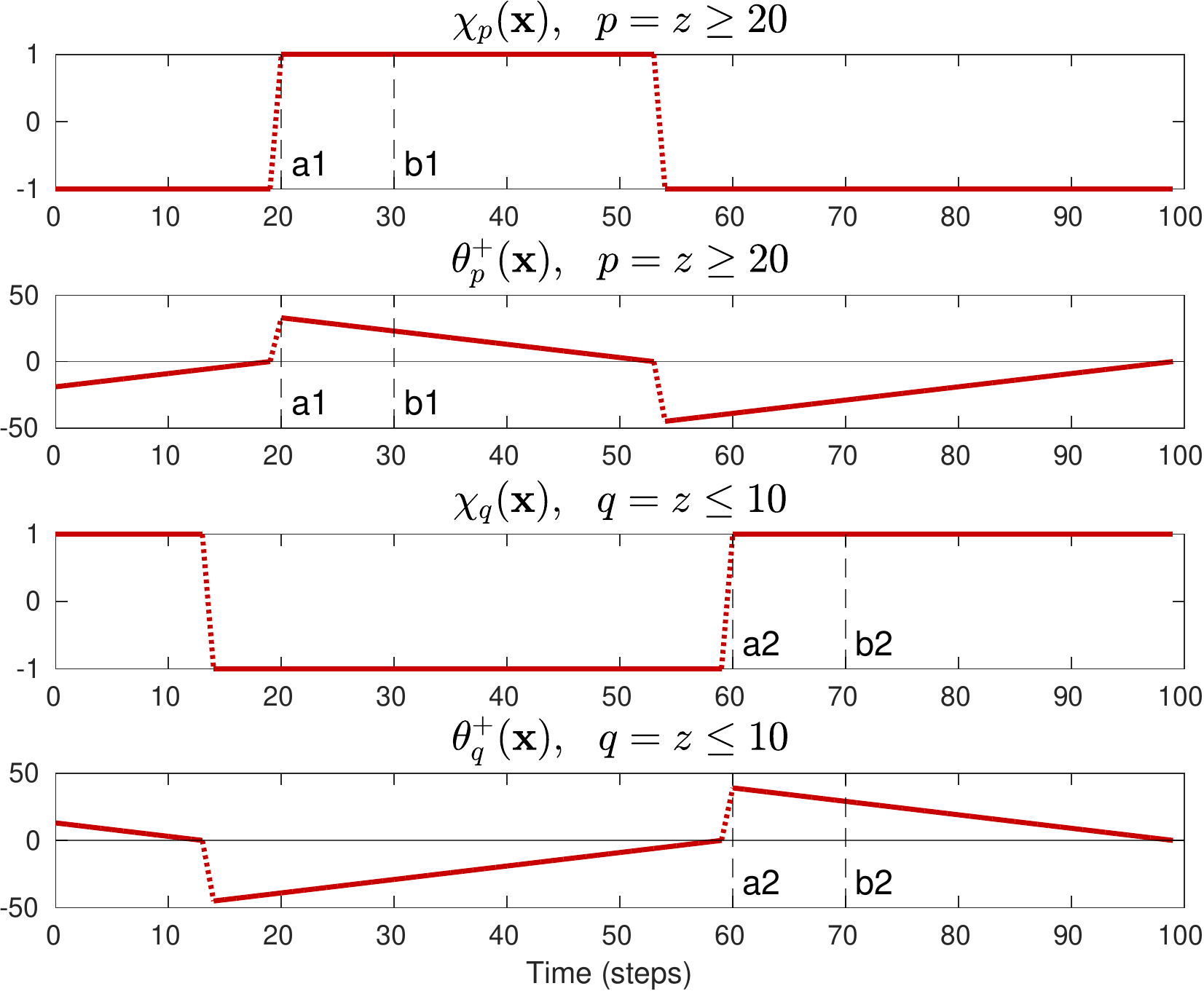}
		\caption{Evolution of the characteristic function $\chi(\sstraj)$ and the right time robustness $\thetap(\sstraj)$ for predicates $p$ and $q$, where $p=z\geq 20$ and $q=z\leq 10$.}
	\end{subfigure}
	\vspace{-5pt}
	\caption{\small UAV Altitude Control. Found maximum right time robustness is $\thetap_{\varphi_{uav}}(\sstraj,0) = 23$ time steps.}
	\label{fig:ex_drone}
	\vspace{-4mm}%
\end{figure*}

\subsection{Case Study 1: UAV Altitude Control}
\label{sec:uav}

Consider a one dimensional unmanned aerial vehicle (UAV) which is moving only in the $z$-axis direction. 
Its state $x=[z, v_z]^\top\in\Re^2$ comprises of altitude $z$ and velocity $v_z$. Initial UAV position and velocity are assumed to be zero. 
The UAV discrete-time linear dynamics are defined as:
\begin{equation}
	x_{t+1} = Ax_{t}+Bu_{t},\quad x_0=[0,0]^\top
\end{equation}
where $A=\begin{bmatrix}
	1&1\\0 &1
\end{bmatrix}$, $B=\begin{bmatrix}
0.5\\1
\end{bmatrix}$ and $|u_t|\leq u_{\max}=0.2$.
In this case study, the UAV is tasked to reach and stay above $z\geq20$ altitude during the time interval $[20,30]$ and then fly down and stay below $z\leq10$ during $[60,70]$ time steps.
The UAV should also maintain its velocity within $|v_t|\leq v_{\max}=1.5$. Time horizon is set to $H=100$. 
The mission is captured in the following specification:\footnote{Since the control input and velocity constrains specified above are of the form $\always_{[0,H]} p$, their maximum time robustness is 0. Therefore, they were removed from the mission specification $\varphi$ and were implemented as strict MILP constraints.}
\begin{equation}
	\label{eq:ex_drone}
	\begin{aligned}
		\varphi_{uav} = \always_{[20,30]} (z \geq 20) 
		\, \wedge\, 
		&\always_{[60,70]} (z \leq 10).
	\end{aligned}
\end{equation}

\textbf{Results.} Solving Prob.~\ref{prob:synthesis} for $\bowtie=+$ gives optimal solution $\thetap_{\varphi_{uav}}(\sstraj,0)=23$, see Fig.~\ref{fig:ex_drone}. 
One can see that even if the UAV started execution of its trajectory earlier by up to 23 time steps, the mission specification would still be satisfied. 
In Fig.~\ref{fig:ex_drone}(a) one can see that UAV indeed stays above $z=20$ from $20$ to $30$ time steps (depicted in green) and then continues being above $z=20$ for the next 23 time steps (depicted in yellow). 
On Fig.~\ref{fig:ex_drone}(b) this is seen as $\chi_{p}(\sstraj,t)=+1$ for all time steps from $20$ to $53$, where $p=z\geq 20$. For these time steps, $\thetap_p(\sstraj,t)$ linearly decreases from $33$ (when $t=20$) first to $23$ ($t=30$) and then to 0 ($t=53$). Which is expected since the right time robustness counts the sequential steps to the future.    
Above explanation applies analogously to the second subpart of the formula $\varphi_{uav}$ when $z\leq 10$. For the computational complexity of this case study see Table~\ref{tab:experiments}.

\subsection{Case Study 2: Multi-agent Surveillance}
\label{sec:ex_agents}

\begin{figure*}[t!]
	\centering
	\begin{subfigure}[t]{.85\columnwidth}
		\centering
		\includegraphics[height=5.5cm]{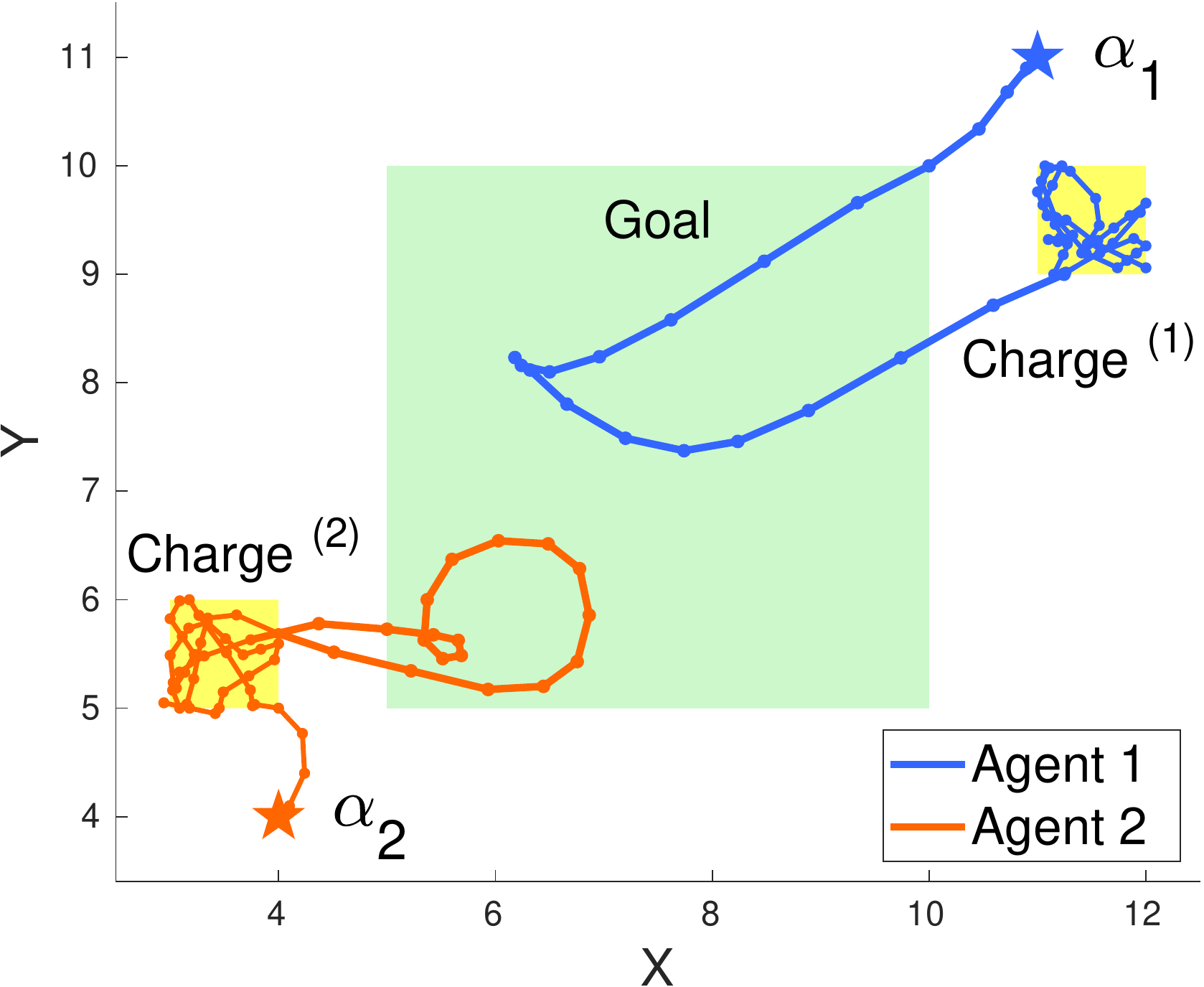}
		\caption{Trajectories for two agents generated by solving Prob.~\ref{prob:synthesis}. Agents perform surveillance and recharging. Initial positions marked by $\star$.}
	\end{subfigure}
	\hspace{15pt}
	\begin{subfigure}[t]{1.1\columnwidth}
		\centering
		\includegraphics[height=5.5cm]{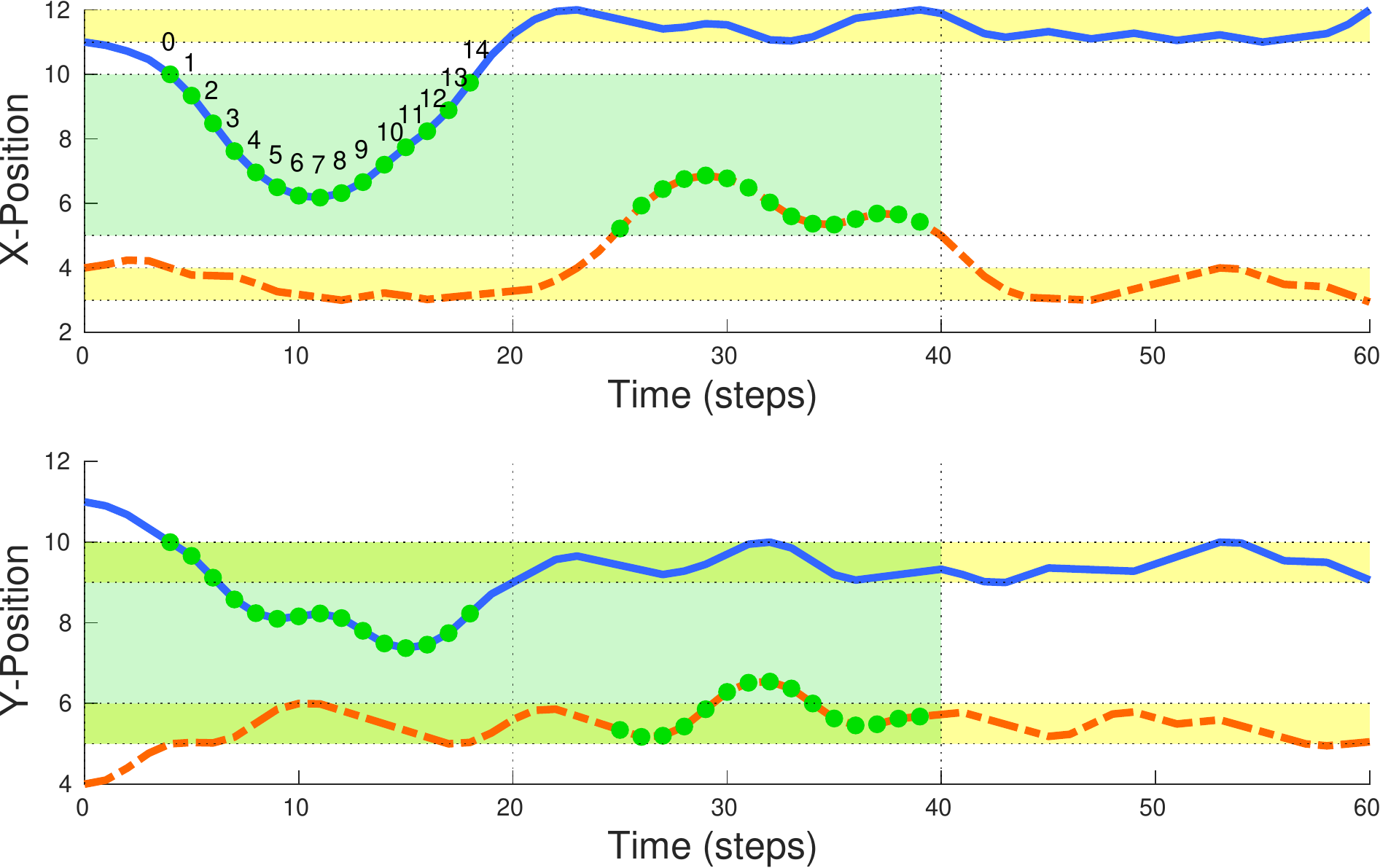}
		\caption{Trajectory projection on X-axis and Y-axis. The found robustness value is $\thetap_{\varphi_{surv}}(\sstraj,0) = \thetap_{\varphi_{gs}}(\sstraj,0) = 14$. We depict the meaning of $\thetap_{\varphi_{gs}}(\sstraj,0)$ as green dots along the trajectory.}
	\end{subfigure}
	\vspace{-6pt}
	\caption{\small Multi-agent surveillance. Goal set is represented in green color, charging zones are in yellow color. Found maximum right time robustness is $\thetap_{\varphi_{surv}}(\sstraj,0) = 14$ time steps. Simulation is available at \href{https://tinyurl.com/multi-surveil}{https://tinyurl.com/multi-surveil}.}
	\label{fig:ex_agents(1)}
	\vspace{-2mm}
\end{figure*}

\begin{figure*}[t]
	\centering
	\includegraphics[width=\textwidth]{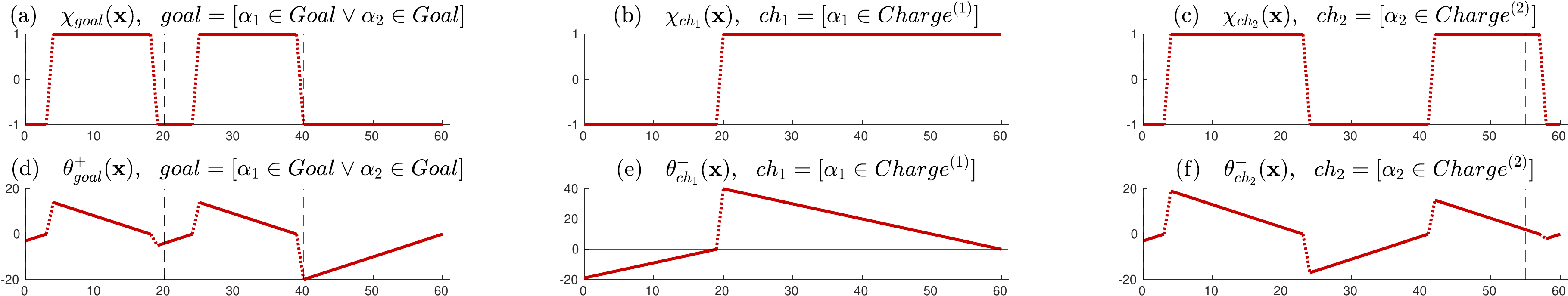}
	\caption{\small Multi-agent surveillance. Evolution of the characteristic function $\chi(\sstraj)$ and the right time robustness $\thetap(\sstraj)$ for sub-formula $goal$ and predicates $ch_1$, $ch_2$. Fig.~\ref{fig:ex_agents(pred)}(a),(d): The \textit{goal} is satisfied during two disjoint time intervals, each interval consists of 14 time steps, $\thetap_{goal}(\sstraj,4)=\thetap_{goal}(\sstraj,25)=14$. Fig.~\ref{fig:ex_agents(pred)}(b),(c),(e),(f): The maximum value of $\thetap(\sstraj)$ for both predicates $ch_1$ and $ch_2$ is higher than 14.}
	\label{fig:ex_agents(pred)}
	\vspace{-6mm}
\end{figure*}

We now formalize the case study with multiple agents carrying out a surveillance mission. 
Consider two identical agents moving in a two-dimensional
space, see Fig.~\ref{fig:ex_agents(1)}(a). They are tasked with a surveillance mission of the region of interest while having finite battery lives and specific re-charging schedules. 
The space consists of two electric agents $\alpha_i$, $i\in\{1,2\}$, one region of interest denoted as $Goal$ and two charging stations $Charge^{(i)}$, $i\in\{1,2\}$. 
Let the state of each agent be
$x^{(i)}=[x^i,\, v^i_x,\, y^i,\, v^i_y]^\top\in\Re^4$ and the control input be $u^{(i)}=[u^i_x,\ u^i_y]^\top\in\Re^2$.
We denote the full state of the system as $x=[x^{(1)},\ x^{(2)}]^\top$ and full control as $u=[u^{(1)},u^{(2)}]^\top$.
The linear state-space representation of the system is given by:
\begin{equation}
	x_{t+1}=Ax_t+B u_t, \quad ||u_t||_\infty\leq 20,
\end{equation}
where
$A=I_4 \otimes \begin{bmatrix}
	1&0.1\\0&1
\end{bmatrix}$ and $B=I_4 \otimes\begin{bmatrix}
0.005\\0.1
\end{bmatrix}$, with $I_n$ being an identity matrix of
dimension $n\times n$ and symbol $\otimes$ denoting the Kronecker product.
Initial positions of the agents are $(11, 11)$ and $(4,4)$, initial velocities are zero. We set time horizon to $H=60$. 

The goal surveillance sub-mission requires the \textit{Goal} region to be visited by at least one of the agents within the first 20 time steps and then within the next 20 time steps. Formally, it is defined as:
\begin{align*}
	\varphi_{gs}= &\eventually_{[0,20]} \left[\alpha_1\in Goal \ \vee\ \alpha_2\in Goal\right] \\
	&\ \ \quad\wedge\  \eventually_{[20,40]} \left[\alpha_1\in Goal \ \vee\ \alpha_2\in Goal\right].
\end{align*}
where notation $\alpha_i\in Z$ denotes more formal $[x^i,\,y^i]^\top \in Z$, i.e. position of the agent $\alpha_i$ is within the given rectangle $Z$. This can be defined as a conjunction of four linear predicates $x^i \leq X_{ub}$, $x^i \geq X_{lb}$, $y^i \leq Y_{ub}$, $y^i \geq Y_{lb}$. 
 
 The battery
 life of the first agent is 20 time units and charging takes 20 time units during which the agent should stay at its charging station. 
 Such pattern must be satisfied at all-time within the given time horizon of the formula: 
 \begin{equation}
 	\varphi_{ch,1}=\always_{[0,20]} \eventually_{[0,20]} \always_{[0,20]}\,  [\alpha_1\in Charge^{(1)}]. 
 \end{equation}

Second agent must satisfy a less restrictive battery charging schedule which is formally specified as:
 \begin{equation}
	\begin{aligned}
		\varphi_{ch,2}= &\eventually_{[0,20]} [\alpha_2\in Charge^{(2)}] \\ &\quad\ \ \wedge\ \eventually_{[40,55]} [\alpha_2\in Charge^{(2)}].
	\end{aligned}
\end{equation}
The overall multi-agent surveillance mission is defined as:
\begin{equation}
	\label{eq:mission2}
	\varphi_{surv} = \varphi_{gs}\, \wedge\, 
	\varphi_{ch,1} \,\wedge\, \varphi_{ch,2}.
\end{equation}


 \textbf{Results.}
 Solving Prob.~\ref{prob:synthesis} for $\bowtie=+$ gives optimal solution $\thetap_{\varphi_{surv}}(\sstraj,0)=14$ time units, see Fig.~\ref{fig:ex_agents(1)} and Fig.~\ref{fig:ex_agents(pred)}. Simulation is available at \href{https://tinyurl.com/multi-surveil}{https://tinyurl.com/multi-surveil}. 
 
 From Fig.~\ref{fig:ex_agents(1)}(b) one can see that Agent 1 surveils the region of interest \textit{Goal} while Agent 2 is in its charging zone $Charge^{(2)}$, and then Agent 1 goes to charging zone $Charge^{(1)}$ while Agent 2 continues to surveil the \textit{Goal}.
 Table~\ref{tab:experiments} shows that this case study is the most computationally heavy and requires 2750 Boolean and 858 integer variables. Solver needs $49.69$ seconds to solve Prob.~\ref{prob:synthesis}. 
 For comparison, we also consider a variation of formula $\varphi_{surv}$ where we require an additional constraint on velocities of both agents:
 \begin{equation}
 	\label{eq:vel_constr}
 	\varphi'_{surv} = \varphi_{surv} \ \wedge\ \always_{[0,H]} (||v||_\infty\leq 4). 
 \end{equation}
 Such additional constraint did not change the final right time robustness value, though did change the trajectories. Interesting observation is, though $\varphi'_{surv}$ has an additional constraint on top of $\varphi_{surv}$, solver takes only $29.08$ seconds to solve it in comparison with $46.69$ for $\varphi_{surv}$.
  
 \begin{figure}[b!]
 	\centering
 	\begin{subfigure}[t]{.47\columnwidth}
 		\includegraphics[width=\textwidth]{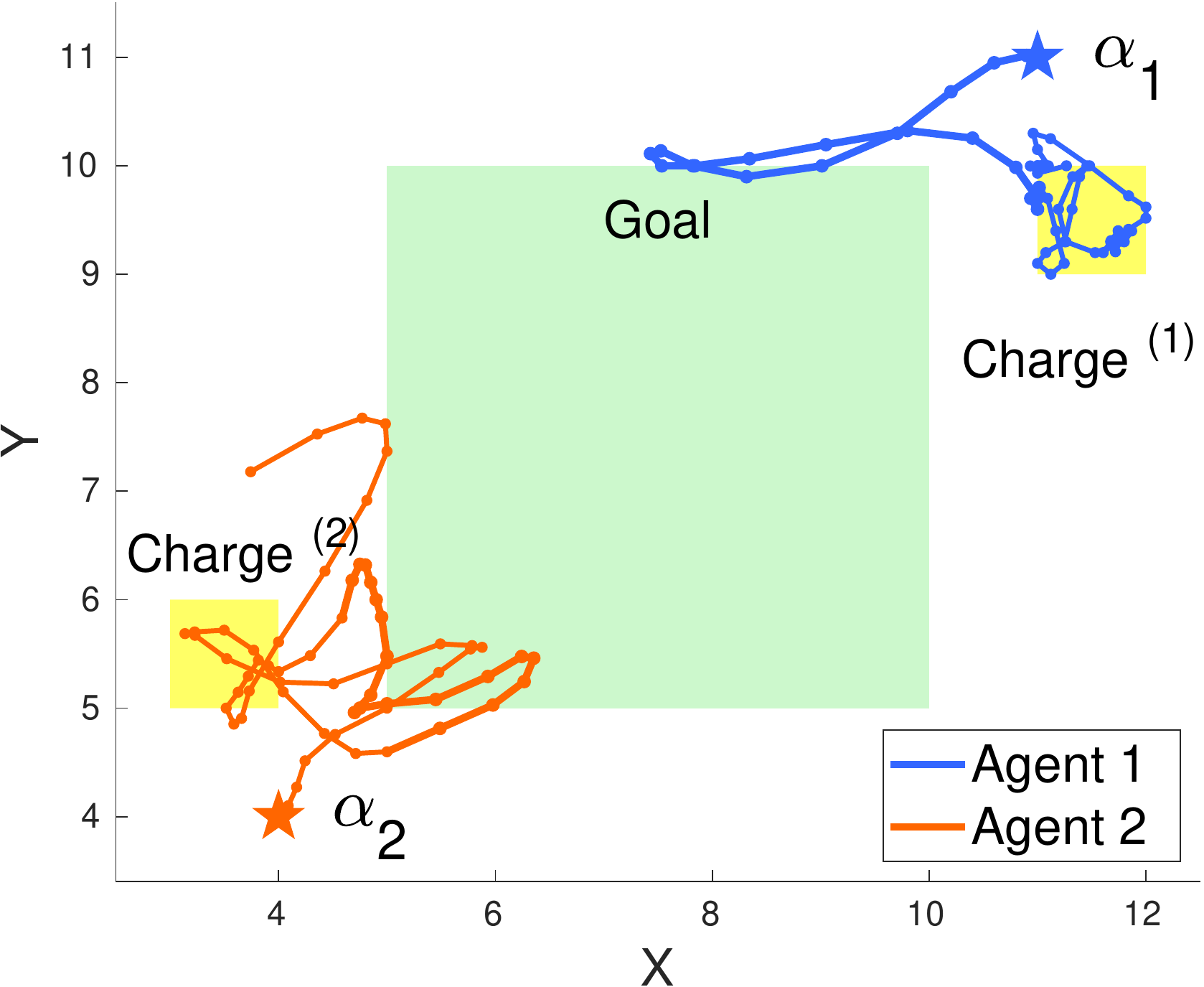}
 		\caption{$\thetap_{\formula_{surv}}(\sstraj,0)=5$.}
 	\end{subfigure}
 	\hspace{5pt}
 	\begin{subfigure}[t]{.47\columnwidth}
 		\includegraphics[width=\textwidth]{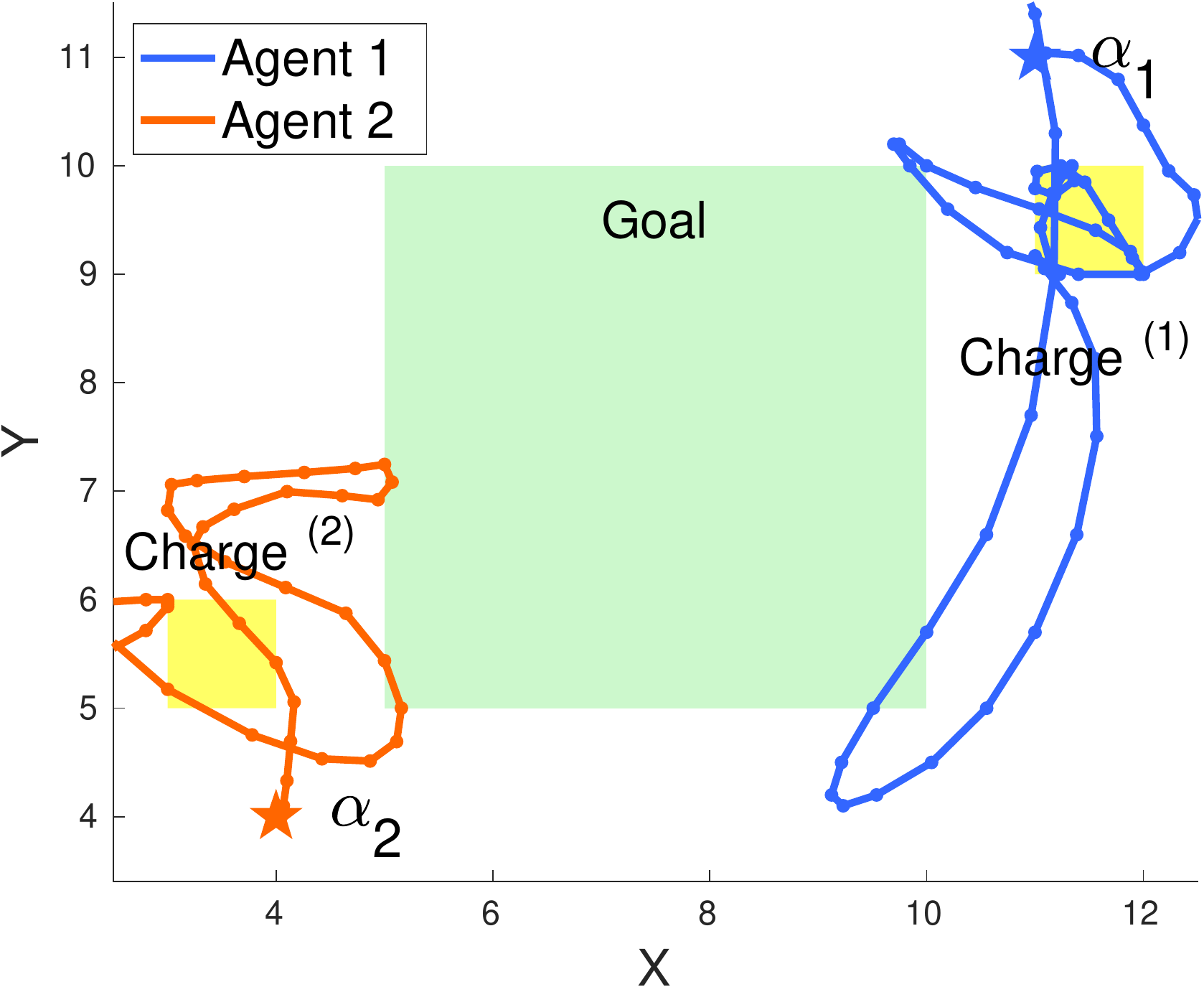}
 		\caption{$\thetap_{\formula_{surv}}(\sstraj,0)=1$.}
 	\end{subfigure}
 	\vspace{-5pt}
 	\caption{\small Solutions to the time robustness feasibility problem instead of the maximization Prob. \ref{prob:synthesis}. Simulations are available at \href{https://tinyurl.com/rob-feasibility}{https://tinyurl.com/rob-feasibility}.}
 	\label{fig:ex_agents(t)}
 \end{figure}

We also performed simulations where we solve a variation of Prob.~\ref{prob:synthesis} where instead of the robustness maximization we solve a feasibility problem that results in robustness having precisely the desired value.
See the resulting trajectories for such feasibility formulation of Prob.~\ref{prob:synthesis} in Fig.~\ref{fig:ex_agents(t)}(a) for $\thetap_{\formula_{surv}}(\sstraj,0)=5$ and  
Fig.~\ref{fig:ex_agents(t)}(b) for $\thetap_{\formula_{surv}}(\sstraj,0)=1$. Table~\ref{tab:experiments} shows that though the size of the problem in terms of number of variables and constraints stays the same, the computation time becomes drastically lower ($2.12$ and $1.02$ seconds).

\section{CONCLUSIONS}
\label{sec:conclusions}

We proposed a controlling system framework for time-critical systems. In particular, we considered system constraints formulated in Signal Temporal Logic (STL). Our framework is based on the Mixed Integer Linear Program (MILP) encoding that allows to maximize the time robustness of STL constraints while guaranteeing to achieve a desired lower bound of the time robustness. We provided correctness guarantees and a complexity analysis of the encoding and illustrated our theoretical findings in two case studies.

\addtolength{\textheight}{-1cm}   




\section*{ACKNOWLEDGMENT}

The authors would like to thank Yash Vardhan Pant for several insightful discussions about time robustness and the control synthesis problem.

\bibliographystyle{unsrt-abbrv}
\bibliography{root}

\section*{APPENDIX}
\subsection{Proof of Theorem \ref{thm:sat}}
\label{sec:proof_sat}

We are going to prove items \ref{i1}) and \ref{i3}) for $\bowtie=+$. Items \ref{i2}) and $\ref{i4})$ for $\bowtie=+$ as well as all the facts $\ref{i1})-\ref{i4})$ for $\bowtie=-$ can be proven analogously. The proof is by induction on the structure of $\varphi$.  

1) $\thetap_\varphi(\sstraj,t) > 0 \Longrightarrow \chi_\varphi(\sstraj,t)= +1$. 

	\begin{itemize}
		\item $\varphi=p$.
		From the definition of $\thetap_p$, since $\tau\geq 0$ then $\chi_p(\sstraj,t) > 0$ and thus, since $\chi\in\{\pm 1\}$, $\chi_p(\sstraj,t)=+1$.
		
		\item $\varphi=\neg \varphi_1$.
		By definition, $\thetap_{\neg \varphi_1}(\sstraj,t) =
		-\thetap_{\varphi_1}(\sstraj,t)> 0$. Therefore, $\thetap_{\varphi_1}(\sstraj,t)< 0$
		and from the induction hypothesis, $\chi_{\varphi_1}(\sstraj,t)=-1$, thus, $\chi_{\neg\varphi_1}(\sstraj,t)=+1$.	
		\item $\varphi=\varphi_1\wedge\varphi_2$.
		Since $\thetap_{\formula_1 \land \formula_2}(\sstraj,t) = \thetap_{\formula_1}(\sstraj,t) \ \sqcap\ \thetap_{\formula_2}(\sstraj,t) > 0$,
		both terms are positive: $\thetap_{\formula_1}(\sstraj,t) > 0$ and $\thetap_{\formula_2}(\sstraj,t) > 0$.
		From the induction hypothesis, $\chi_{\formula_1}(\sstraj,t)=\chi_{\formula_2}(\sstraj,t)=+1$ and thus, $\chi_{\formula}(\sstraj,t)= \chi_{\formula_1}(\sstraj,t) \sqcap \chi_{\formula_2}(\sstraj,t)=+1$.
		\item $\varphi=\varphi_1 \until_I\varphi_2$. 
		Due to the maximum operator in the definition of Until operator, $\exists t'\in t+I$ such that $\thetap_{\formula_2}(\sstraj,t') \ \sqcap \
		\bigsqcap_{t'' \in [t,t')} \thetap_{\formula_1}(\sstraj,t'')>0$.
		Now due to the minimum operators, $\thetap_{\formula_2}(\sstraj,t') >0$ and $\forall t'' \in [t,t')$, $\thetap_{\formula_1}(\sstraj,t'')>0$.
		Therefore, from the induction hypothesis, $\exists t'\in t+I$, $\chi_{\formula_2}(\sstraj,t')=+1$ and 
		$\forall t'' \in [t,t')$, $\chi_{\formula_1}(\sstraj,t'')=+1$. And thus, 
		$\chi_{\formula}(\sstraj,t) = +1$.
		This concludes the proof.
	\end{itemize}

3)  $\chi_\varphi(\sstraj,t)= +1 \quad \Longrightarrow\quad
	\thetap_\varphi(\sstraj,t) \geq 0$. 
	\begin{itemize}
		\item $\varphi=p$. $\chi_p(\sstraj,t) = +1$ thus, by def., $\thetap_p(\sstraj,t)\geq 0$ .
		\item $\varphi=\neg \varphi_1$.
		Since, $\chi_{\neg\varphi_1}(\sstraj,t) = -\chi_{\varphi_1}(\sstraj,t) = +1$, therefore, $\chi_{\varphi_1}(\sstraj,t) = -1$, from the induction hypothesis $\thetap_{\varphi_1} \leq 0$. Since $\thetap_{\neg\varphi_1}(\sstraj,t) = - \thetap_{\varphi_1}(\sstraj,t) \geq 0$. 
		\item $\varphi=\varphi_1\wedge\varphi_2$. Since $\chi_\varphi=\chi_{\formula_1} \sqcap \chi_{\formula_2}(\sstraj,t) = +1$ then both $\chi_{\formula_1}(\sstraj,t)=+1$ and $\chi_{\formula_2}(\sstraj,t)=+1$. From induction hypothesis $\thetap_{\formula_1}(\sstraj,t)\geq 0$ and $\thetap_{\formula_2}(\sstraj,t)\geq 0$ and thus, 
		$\thetap_\formula(\sstraj,t)= \thetap_{\formula_1}(\sstraj,t)\sqcap \thetap_{\formula_2}(\sstraj,t) \geq 0$.	
		\item $\varphi=\varphi_1 \until_I\varphi_2$. From the definition of the characteristic function for Until operator,  $\exists t'\in t+I$ such that $\chi_{\formula_2}(\sstraj,t')=+1$ and $\forall t'' \in [t,t')$, $\chi_{\formula_1}(\sstraj,t'') = +1$. By the induction hypothesis we obtain that $\thetap_{\formula_1}(\sstraj,t')\geq 0$ and
		$\thetap_{\formula_1}(\sstraj,t'')\geq 0$
		and thus we conclude that
		$\thetap_\formula(\sstraj,t)\geq 0.$
		This concludes the proof.
	\end{itemize}

\end{document}